\documentclass[submit]{smj}

\usepackage{multirow}
\usepackage{placeins}
\usepackage{amsfonts}
\usepackage{mathbbol}
\usepackage{array}
\newcolumntype{L}[1]{>{\raggedright\let\newline\\\arraybackslash\hspace{0pt}}m{#1}}
\newcolumntype{C}[1]{>{\centering\let\newline\\\arraybackslash\hspace{0pt}}m{#1}}
\newcolumntype{R}[1]{>{\raggedleft\let\newline\\\arraybackslash\hspace{0pt}}m{#1}}

\Author{Murilo S. Pinheiro, 
        Aluísio Pinheiro and Benilton S. de Carvalho, 
}
\AuthorRunning{Pinheiro et al. \textrm{et al.}}

\Affiliations{

\item University of Campinas, Brazil

}   

\CorrAddress{Aluísio Pinheiro, 
             Department of Statistics, 
             IMECC University of Campinas,
             Rua Sérgio Buarque de Holanda, 651 13083-855, 
             Sokolovsk\'a 83, 
             D. Barão Geraldo, Campinas, SP, Brazil.}
\CorrEmail{pinheiro@ime.unicamp.br}
\CorrPhone{(+55)(19)3521-6080}
\CorrFax{(+55)(19)3521-6080}

\Title{A CUSUM approach to the detection of copy-number neutral loss of heterozygosity}
\TitleRunning{Template paper}

\Abstract{
Several genetic alterations are involved in the genesis and development of cancers. The determination of whether and how each genetic alterations contributes to cancer development is fundamental for a complete understanding of the human cancer etiology. Loss of heterozygosity (LOH) is one of such genetic phenomenon linked to a variate of diseases and characterized by the change from heterozygosity (the presence of both alleles of a gene) to to homozygosity (presence of only one type of allele) in a particular DNA locus. Thus identification of DNA regions where LOH has taken place is a important issue in the health sciences. In this article we formulate the LOH detection as the identification of change-points  in the parameters of a mixture model and present a detection algorithm based on the cumulative sums (CUSUM) method. We found that even under mild contamination our proposal is a fast and reliable method.}


\begin{document}

\maketitle

\section{Background}
Several genetic alterations such as single base substitution, translocation, copy-number alteration (CNA) and loss of heterozygosity (LOH) are involved in the genesis and development of cancers \citep{Albertson,Beroukhim,Stratton}. Determining whether and how genetic alterations contribute to cancer development is paramount for human cancer etiology. One of the most common genotyping tool for the identification of those altered regions are single nucleotide polymorphism arrays (SNP Arrays). With resolution up to one marker for every 100 bp SNP Arrays greatly increased the ability of geneticists to explore the structure of DNA and its effect on health and disease. Although recent technology offers even greater resolution the cost of SNP-A makes it a widely used technology to this day.

Copy-number alteration is a type of structural variation in which a particular region of DNA has a number of copies that differs from the expected two in the diploid genome. These alterations can be of inherited origin or the consequence of somatic mutations occurring during the development of a tumor. Although copy-number variations not always malignant, they are a key event in the development of a variety of human cancers. Their role in tumor development is not clear, but increased copy-number of regions containing oncogenes and the deletion of tumor suppressor genes have been widely documented \citep{Beroukhim}.

The primary interest of this text relates to loss of heterozygosity (LOH). This alteration refers to a change from heterozygosity (the presence of both alleles of a gene) to to homozygosity (presence of only one type of allele) in a particular DNA region. LOH can result from the a deletion in a heterozygous DNA region or more intricate phenomena such as mitotic recombination or non-disjunction in somatic tumor cells \citep{Teh}. This latter case of LOH is called copy-number neutral loss of heterozygosity (CNNLOH) and gained increased attention from geneticists since SNP-A made possible their detection in unpaired tumor samples \citep{Bignell,Huang}. LOH in tumor development is often associated to the inactivation of tumor suppressor genes \citep{Janne}.

The development of SNP Arrays technology made it possible for geneticists to look for CNAs and LOH regions in paired and unpaired tumor samples as described below. Two measurements are obtained from such experiments: the first one, called log R ratio (LRR), is given by
$$LRR = \log_2 \frac{R_\text{observed}}{R_\textbf{expected}},$$
where $R_\text{observed}$ is the sum of observed measures for each possible allele and $R_\textbf{expected}$ is the expected sum value. The second one, B allele frequency (BAF)
$$
BAF = \left\{
\begin{array}{ll}
0& \text{if }\theta < \theta_{AA}\\
0.5(\theta - \theta_{AA})/(\theta_{AB}-\theta_{AA})& \text{if }\theta_{AA}\theta < \theta_{AB}\\
0.5 + 0.5(\theta - \theta_{AB})/(\theta_{BB}-\theta_{AB})&  \text{if }\theta_{AB}\leq\theta < \theta_{BB}\\
1&\text{if }\leq\theta \geq \theta_{BB} \\
\end{array} 
\right.
$$ where $\theta = \arctan(X_1/X_2)/(\pi/2)$ is a measure of relative allele frequency. It is clear from those definitions that LRR is related to the copy-number of a locus, while BAF is relative to one of the alleles proportion. 

\begin{figure}[h]
	\centering
	\includegraphics[scale=0.45]{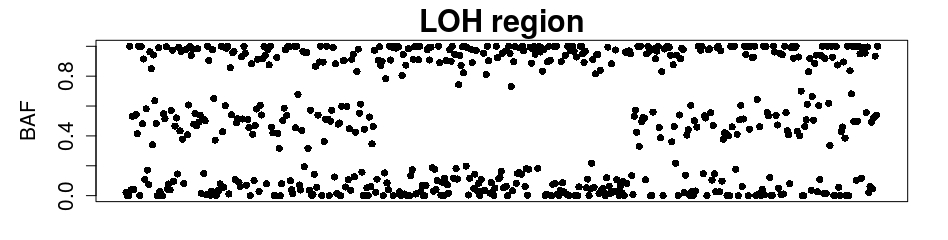}
	\caption{\label{LOHregion} BAF sequencing where the central region is affected by LOH. We can see that what characterizes the presence of LOH is the persistent absence of observations close to 0.5.}
\end{figure}

We present a new method for detecting CNNLOH regions based exclusively in the BAF sequence. Segments of the DNA where LOH has taken place are detected in the BAF plot as the absence of a central band (or bands in the case of copy-number four), as is illustrated in \autoref{LOHregion}.  

Several approaches have been proposed for the identification of CNA and LOH in paired and unpaired comparative genomic hybridization and SNP Arrays data, such as agglomerative clustering \citep{Wang}, penalized likelihood \citep{Picard}, circular binary segmentation \citep{CBS}, piecewise linear models \citep{Muggeo} and hidden Markov models (HMM) \citep{oncoSNP,pennCNV,Beroukhim2006, Ha, dchip}. The HMM-based approach is by far the most widely adopted. Our approach to LOH will be one of identifying changes in the proportion parameters of a mixture model by means of statistical process control (SPC) tools and, thus, without imposing any restriction on the statistical distribution of change-points.

\section{Method}
\subsection{Statistical Model}\label{model}
We assume that $\{x_1,\ldots,x_n\}$ is the BAF data resulting from a SNP Arrays study and that we already know the copy number in this sequence is two. We are interested in detecting regions of this sample where the BAF data appears to be lacking the central band. This means that differences within the upper and lower strips are irrelevant for our purposes. For this reason we perform the following transformation on the data: $y_i = 2\times|x_i-1/2|$. This transformed sample $\{y_1,\ldots,y_n\}$ will be called transformed BAF (tBAF).
\begin{figure}[h]
	\centering
	\includegraphics[scale=0.7]{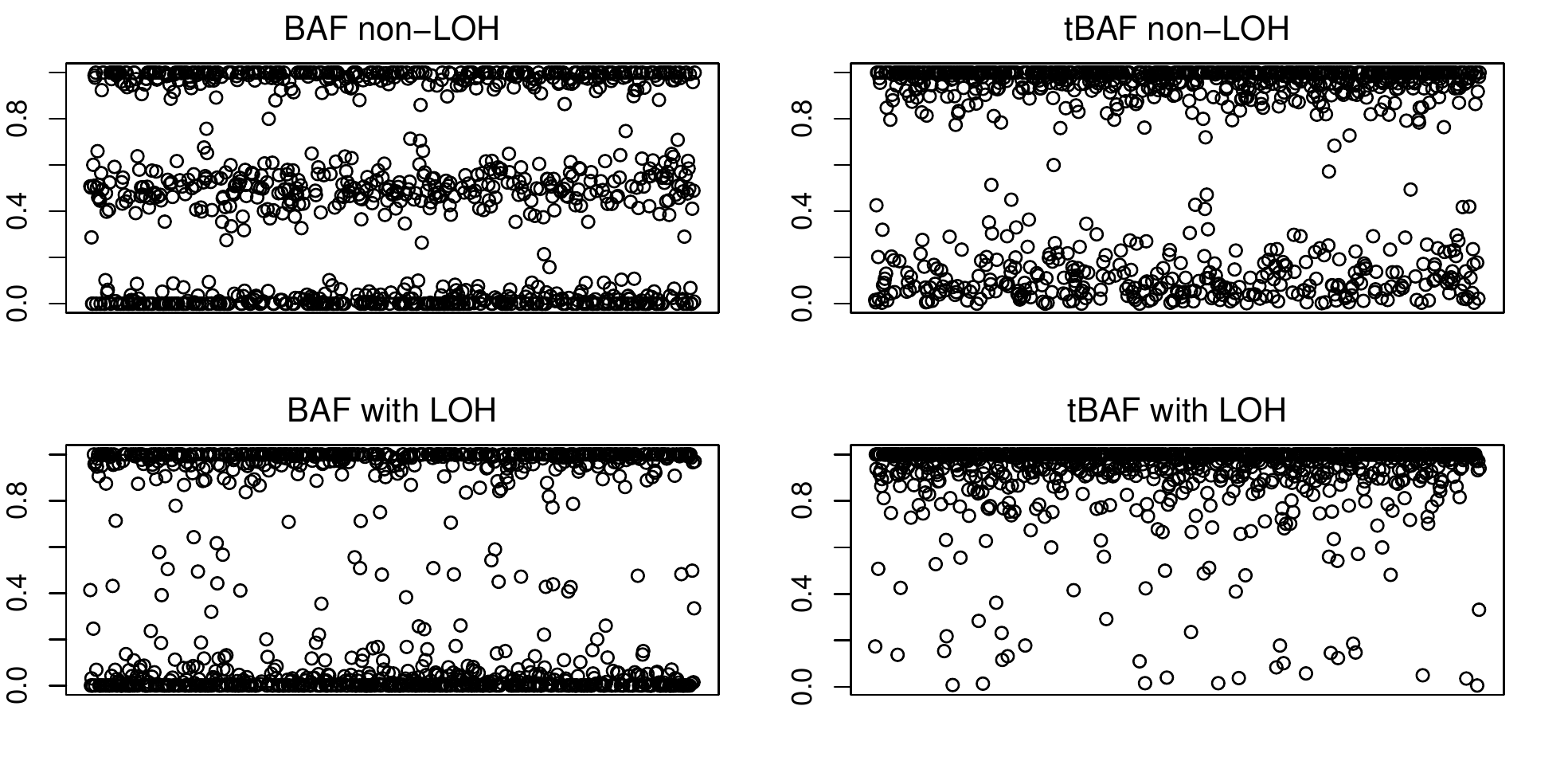}
	\caption{\label{BAFtBAF}BAF and tBAF plots for LOH and non-LOH regions. Considering first the non-LOH region we can see that after transformation the TBAF sequence is characterized by two observation strips, one running close to 1 and another close to 0. When a region affected by LOH is transformed there are very few observations close to 0 but the strip close to 1 is still present.}
\end{figure}

The tBAF data follows a pattern illustrated in \autoref{BAFtBAF}. It can be seen  that for non-LOH regions there are clearly two bands in the tBAF plot, one running near $1$ and another near $0$. We will adopt the convention of calling {\it upper band} the band that runs near $1$ and {\it lower band} any band running below the upper one. The presence of LOH can be identified in the tBAF plot by the absence of an identifiable lower band.

We propose adopting the mixture of two distributions as a model for the tBAF distribution. The first distribution describes the the upper band stochastic behavior whilst the second component should yield the lower band behavior. Take $f_0$ as the density associated to observations in the lower band and $f_1$ for the observations on the upper band. Our model for both non-LOH and LOH regions will be of the form:
$$p(\boldsymbol y|\boldsymbol\xi) = (1-\pi) f_0(\boldsymbol y|\boldsymbol\xi) + \pi f_1(\boldsymbol y|\boldsymbol\xi)$$
where $\boldsymbol y$ is the tBAF vector of observations, $\boldsymbol\xi$ is a vector containing all the necessary parameters for characterizing our distribution and $\pi$ is the probability of drawing a observation from $f_1$. The parameter $\pi$ can be interpreted as the probability of observing a heterozygous DNA locus and will be called here the {\it homozygosity level}. In some methods it is assumed known \citep{chen} but its availability for a particular combination of platform and population is not always warranted. For this reason we propose an estimation method that requires only the available sample. 

We have argued that what characterizes a LOH region is the lack of any lower band. This can be rephrased as a difference in the parameter $\pi$ in our statistical model, i.e., we expect that the density function describing LOH regions have a much smaller coefficient associated to observing a value drawn from $f_0$. From this observation we can formulate two such models:
$$p_0(\boldsymbol y|\boldsymbol \xi) = \pi_0f_0(\boldsymbol y|\boldsymbol\xi) + (1-\pi_0)f_1(\boldsymbol y|\boldsymbol\xi),$$
$$p_1(\boldsymbol y|\boldsymbol\xi) = \pi_1f_0(\boldsymbol y|\boldsymbol\xi) + (1-\pi_1)f_1(\boldsymbol y|\boldsymbol\xi),$$
where $p_0$ is associated to non-LOH regions, $p_1$ to LOH regions and $\pi_0 > \pi_1$. In fact, since the LOH is supposed to lack it's lower band, we may assume that $\pi_0 = \delta\pi_1$ with $0\leq \delta<1$.

We adopt the one inflated beta (OIB) density function \citep{inflated}, which has the form:
$$f_{OIB}(y|\theta_1,\alpha) = \left\{\begin{array}{lll}
\theta_1, & \text{if} & y = 1 \\ 
(1-\theta_1)\alpha y^{\alpha - 1}, & \text{if} & 0<y\leq1
\end{array}\right.,  $$
where $\theta_1\in[0,1]$ is the probability of observing a $1$ and $\alpha\in(0,+\infty)$.

The choice of $f_0$, which describes the lower band, is the zero inflated beta (ZIB) distribution \citep{inflated}, whose density is given by: 
$$f_{ZIB}(y|\theta_0,\beta) = \left\{\begin{array}{lll}
\theta_0 & \text{if} & y = 0 \\ 
(1-\theta_0)\beta (1-y)^{\beta - 1} & \text{if} & 0\leq y<1
\end{array}\right.  $$
where $\theta_0\in[0,1]$ is the probability of observing a $0$ and $\beta\in(0,+\infty)$.

The estimation of the parameters in the proposed model is performed by Expectation Maximization (EM) algorithm \citep{givens} application to a set of observations where CNNLOH is absent. This set can be obtained from the application of the microarray technology to somatic tissue or by visually inspecting the tBAF sequence and identifying regions without LOH.

\subsection{LOH Calling}
As discussed in \autoref{model} the detection of LOH regions in a tBAF sequence can be formulated in the framework of statistical process control (SPC), i.e., the change from non-LOR, in control, to LOH, out of control, or vice-versa. Despite this connection there has been little work on the application of this perspective to the problem of segmenting SNP Arrays data, the work of Li et al. \citep{Li} being the only example known to the authors.

There are a number of possible methods to perform SPC in a sequence of observations \citep{BasseNiki1993}. The CUSUM method, which was first proposed by Page in \cite{Page1}, is one of the less commonly adopted \citep{Hawkins}. The reason is that the CUSUM plots are especially designed to identify a change in a parameter value (or a distribution), to a known value (or distribution) after a change. This assumption of knowledge regarding prior and after change parameters values is usually not reasonable in real applications \citep{BasseNiki1993}.

Although usually seen as a disadvantage this characteristic of the CUSUM plot is taken here as a advantage. As was shown in \autoref{model}, we are able to provide good approximations for in-control and out-of-control distributions previous to the application of any SPC tool. The CUSUM is not only specially adapted to the situation of a change in known distributions but was shown to be optimal \citep{CUSUMoptimal} and also asymptotically optimal \cite{Lorden} if one sets the average delay to detection minimization as the optimality criterion, i.e., it is (in some sense) the fastest method to identify a change-point.

\subsection{CUSUM Change-point Detection Procedure}
We now provide a detailed account of the CUSUM algorithm for our application. For  notational convenience we will always assume that a transition of $p_0$ to $p_1$ is to be detected.

We suppose known whether the actual segment is of the non-LOH type. This means that we are working under the assumption that the observations are independent realizations of $p_0$. We say that $p_0$ is the {\it assumed model}. To detect a transition from $p_0$ to $p_1$ the CUSUM algorithm sequentially computes the instantaneous log-likelihood: $s_i = \log p_1(y_i) - \log p_0(y_i)$ and its cumulative sums: $S_0 = 0$; $S_i = \max\{0,S_{i-1} + s_i\}\;\;i\geq 1$.

The presence of a change-point is flagged, at time $t$, if $S_t$ is the first cumulative sum to be greater than the alarm threshold $L_0$.

When the CUSUM detects the presence of a change-point the next step is to estimate its location. We adopt a maximum likelihood approach to solve this problem. Let the supposed model be $p_0$, that we began the cumulative sum at $t = i_1$ and that the alarm time was $t = i_2$. Our estimate of the change-point $\tau$ is:
$$\hat{\tau} = \underset{i_1 \leq t \leq i_2}{\text{argmax}}\left\{\sum_{j = i_1}^{t-1}\log p_0(y_j) + \sum_{j = t}^{i_2}\log p_1(y_j)\right\} $$

Once the change-point is estimated, all observations between the previous change-point and the newly discovered one are said to be realizations of $p_0$ (non-LOH) and the CUSUM algorithm restarts at the last change-point now considering $p_1$ as the assumed model. When the assumed model is $p_1$ the CUSUM statistic changes to $s_i = \log p_0(y_i) - \log p_1(y_i)$ and the threshold to $L_1$. A similar procedure is then carried out until the next change-point comes by and we return to $p_0$ after identifying a LOH region. 

The main issue with the CUSUM algorithm is the correct choice of the two alarm thresholds. Usually those thresholds are found based on the average run length functions $ARL^i_j(L_i)$ for a threshold value $L_i$, for $i,j\in\{0,1\}$. The $i$ index is that of the assumed model and $j$ that of the observed model, i.e., if $i = 0$ and $j = 1$ the assumed model is $p_0$ and the observations are been generated from $p_1$. It follows that, for example, $ARL^0_1$ is the expected time to call a change-points from $p_0$ to $p_1$ and $ARL^0_0$ is the amount of time to a false alarm when the assumed model is $p_0$.

\cite{Page1} shows that the ARL function is the solution for an integral equation of the Fredholm's type but an analytical solution does not always exist. To overcame this difficulty numerical approximations have been suggested \citep{Goel,Page2,Sieg,Brook,Lucas1}. In the next subsection we propose a novel approach to the threshold selection.

\subsection{Threshold Approximation}
Here we do not base threshold values on the previous approximations of the ARL function. The choice of a particular value of average run length is difficult and it has no theoretical meaning to geneticists performing the analysis. We propose a criterion for selecting a threshold based on the a definition of segments with small length and a level of tolerance for segments with smaller lengths as follows. 

We assume that a index can only be called a change-point if the following segment has length at least $m$. This restriction of a minimum length can be integrated in the CUSUM method by imposing a condition on the probability of raising an alarm only after changes that persist for at least $m$ observations. Take $p_0$ as our supposed model and let $\{y_1,y_2,\ldots\}$ be a sequence of independent realizations of $p_1$. Consider $S_i$, $i = 0,\,1,\ldots$, the sequence of sums resulting from an application of the CUSUM algorithm to $\{y_1,y_2,\ldots\}$. Given a threshold $L$ and a probability $\alpha\in(0,1)$, the previous restriction on the minimum length of a segment can be formulated as:
\begin{equation}\label{lim}
\mathbf{P}(R_m < L) \geq 1 - \alpha
\end{equation}
where $R_m = \underset{1\leq i \leq m}{\max}S_i$ and $\alpha$ is the predefined level of tolerance. This restriction can be interpreted as imposing the condition of only raising an early alarm when there is abundant evidence of a change-point.

For any two given values of $m$ and $\alpha$ there are infinitely many values of $L$ such that \autoref{lim} is valid. One of these values is the $(1-p)$th quantile of the $R_m$ statistical distribution. This value of $L$ can be estimated by bootstrapping the distribution of $R_m$ from simulations of the estimated distribution $p_1$. The threshold for the assumed model $p_1$ can be found in a similar way.

\section{Computational Studies}
In this section we present a set of computational simulations where we evaluated the ability of the presented segmentation method to correctly identify regions with and without LOH along a BAF sequence. In order for the study to be as close as possible to the situations encountered in practice we will use the BAF samples available at the ''acnr'' R package \cite{acnr}.

We simulate a set of sequences of the form $ \{X_1, \ldots, X_{1000}\}$ by randomly sampling with replacement observations available in the ''acnr'' package. The structure of the sequences are always as follows: we have $ \{X_1, \ldots, X_ {500}\} $ selected from the population with neutral number of copies and without LOH, then $\{X_ {501},\ldots, X_{501 + l-1}\}$ are selected from the population with CNNLOH and finally $\{X_{501 + l},\ldots, X_{1000}\}$ are sampled from the population with copy-number neutral and without LOH. We use in our study the values of $l \in \{25,50,100\}$. In addition, we consider the cases where the purity of the sample can assume the values $ p = 1, \; 0.79, \; 0.5 $.

In our algorithm we always take $\delta = 10^{-2}$ and $\alpha = 0.05$ as the parameter for LOH model and threshold calculation, respectively. We also consider $m \in \{10,25,50\}$.

In order to evaluate the segmentation quality, we note that the problem of detecting LOH regions can be seen as a classification problem with only two categories. In our case we call negative (0) observations in regions without LOH and positive (1) the observations in regions with LOH. Then for each simulated sequence we have a base sequence $\{t_1, \ldots, t_{1000}\}$ of zeros and ones describing to which class each observation belongs. If $\{I_1,\ldots,I_{1000}\}$ is the sequence of zeros and ones indicating the classification using the proposed method, we define
$$\text{TP} = \sum_{i=1}^{1000}\mathbb{1}\{t_i = 1,\;I_i = 1\},\;\text{FP} = \sum_{i=1}^{1000}\mathbb{1}\{t_i = 0,\;I_i = 1\}$$
$$\text{TN} = \sum_{i=1}^{1000}\mathbb{1}\{t_i = 0,\;I_i = 0\},\;\text{FN} = \sum_{i=1}^{1000}\mathbb{1}\{t_i = 1,\;I_i = 0\}$$

We use the measures of sensitivity $ = TP / (TP + FN) $ and of specificity $ = TN / (TN + FP) $ to evaluate the performance of our method. \autoref{sens} and \autoref{esp} present the means of specificity and sensitivity for 100 replicates of each sequence $\{X_1, \ldots, X_{1000}\}$, respectively.
\begin{table}[htb]
	\caption[Resultados de sensitividade]{\label{sens}Sensitivity results. We can see that when $m$ is small and purity is close or bigger to $0.78$ our method is able to detect a large percentage of the LOH regions. The method's ability to detect LOH regions decreases with the sample purity. }
	\centering
	\begin{tabular}{cc|c|c|c}
		
		&&m = 10 & m = 25 & m = 50\\\hline
		\multirow{3}{*}{Purity = 1}&l = 25&0.97& 0.69& 0\\
		&l = 50&0.98& 0.98& 0.64\\
		&l = 100&0.99& 0.99& 0.99\\\hline
		\multirow{3}{*}{Purity = 0.79}&l = 25& 0.94& 0.33& 0.0000\\
		&l = 50& 0.97& 0.97& 0.06\\
		&l = 100& 0.99& 0.9& 0.90\\\hline
		\multirow{3}{*}{Purity = 0.5}&l = 25& 0.04& 0.01&    0\\
		&l = 50& 0.05&    0&    0\\
		&l = 100&0.07&    0&    0\\
	\end{tabular} 
\end{table}
\begin{table}[htb]
	\caption[Resultados de especificidade]{\label{esp}Specificity results. As one would expect the smaller $m$ is the less specific the method is since smaller values of $m$ make the segmentation more susceptible to false-discoveries. In no instance the false-discovery rate is big enough to cause concerns.}
	\label{espe}
	\centering
	\begin{tabular}{cc|c|c|c}
		
		&&m = 10 & m = 25 & m = 50\\\hline
		\multirow{3}{*}{purity = 1}&l = 25&0.94& 0.99& 1\\
		&l = 50&0.95& 0.99& 0.99	\\
		&l = 100&0.95& 0.99 & 0.99\\\hline
		\multirow{3}{*}{purity = 0.79}&l = 25&0.95& 1& 1\\
		&l = 50& 0.95& 1& 1\\
		&l = 100&  0.94& 0.99& 1\\\hline
		\multirow{3}{*}{purity = 0.5}&l = 25&0.95& 1& 1\\
		&l = 50& 0.95& 1& 1\\
		&l = 100&0.95& 1& 1
	\end{tabular} 
\end{table}

Note that in \autoref{sens} that our method correctively detected regions with LOH in cases with purity equal to one and $m<l$. When $m \leq l$ the segmentation quality is noticeably lower. As the purity decreases the segmentation quality also becomes worse. For the case of purity 0.5 the method was unable to correctly identify the LOH regions in the vast majority of cases. For 0.79 purity the situation is not as severe with the case $m = 10$, which yields reasonably good results.
 
\autoref{esp} shows that our proposed method does not overestimate the regions with LOH but miss some of then when the value of $m$ is close or bigger than the region length or when the sample purity is close to 0.5.

\section{Real Data Application}
We will now apply the proposed method to a real data set. The data set we utilize consists of 482 benign and tumor samples from 259 men with prostate cancer studied in \cite{sampleref}.

We apply the oncoSNP segmentation procedure \citep{oncoSNP} to all tumoral samples available in ''https://www.ncbi.nlm.nih.gov/geo/'' and separate all the segments where the method accused a neutral number of copies. In each of these segments we apply our segmentation method and compare the detection of regions with LOH to the one made by oncoSNP again using sensitivity and specificity assuming that the result of oncoSNP is the gold standard.

We visually select a segment with 1800 observations within chromosome 1 and estimate our models using this segment. The other parameters of our method were chosen as $\alpha = 0.05 $, $\delta = 0.01$ and we used 10000 simulations to estimate the two threshold values. \autoref{comp} presents the sensitivity and specificity results.

\begin{table}[htb]
	\caption{\label{comp} Mean execution time, sensitivity and specificity of our method assuming the results of oncoSNP as the gold standard.}
	\label{sens_onco}
	\centering
	\begin{tabular}{c|c|c|c|c}
		
		& m = 25&m = 50 & m = 100 & m = 150\\\hline
		Execution time (s) & 46.6212 & 47.2890 & 46.2117 & 44.6742\\
		Sensitivity&0.9517& 0.9050& 0.8269& 0.7958\\
		Specificity&0.7939& 0.8706& 0.9392& 0.9573
	\end{tabular} 
\end{table}

It is clear that for all values of $m$ our method correctly detected most of the LOH regions pointed out by oncoSNP. As one would expect smaller values of $m$ detect more oncoSNP segments than do greater values of $m$. In terms of specificity the opposite is true: greater values of $m$ result in greater specificity. This is also not surprising.

In terms of execution time our procedure shows a great advantage in comparison to oncoSNP. The worst mean execution time of our procedure is 47.29 seconds. Also if we remove the time necessary for data loading this times reduces to,a maximum of, 9.72 seconds. This is a very small execution time when compared to the minuties mean execution time
for oncoSNP.

\begin{figure}[h]
	\caption[Perfiz de segmentação]{\label{comparizon_CUSUM}Segmentation profiles for (a) oncoSNP and for different parametrization of our method (b) $m = 100$ e $\delta = 10^{-2}$, (c) $m = 50$ e $\delta = 10^{-2}$ e (d) $m = 50$ e $\delta = 10^{-4}$. The red lines indicates a transition between regions.}
	\begin{center}
		\includegraphics[scale = 0.8, angle=-90]{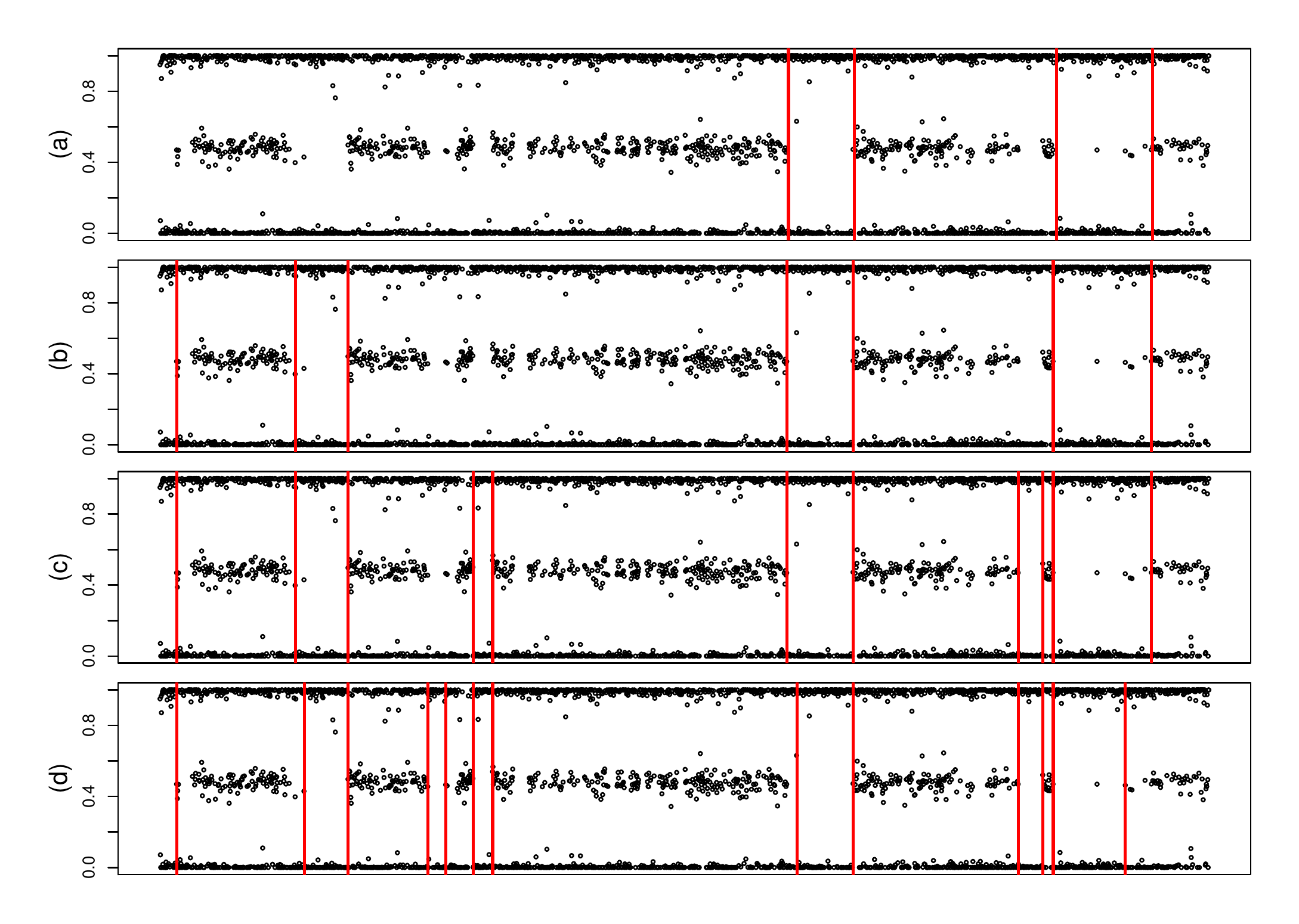}
	\end{center}
\end{figure}

To explore the segmentation features we choose a small region of one of the segments and look at the regions of LOH detected by oncoSNP and by our method. We look at the segmentation made by three different parameter choices for our method:  $m = 100$ and $\delta = 10^{-2}$, $m = 50$ and $\delta = 10^{-2}$, $m = 50$ and $\delta = 10^{-6}$. In all cases we adopt $\alpha = 0.05 $ and 10000 simulations in the thresholds estimation process. The segmentation results are presented in \autoref{comparizon_CUSUM} where the order of annotation of the graphs (a), (b), (c) and (d) follows the order in which we presented the algorithms.

The first thing we note in \autoref{comparizon_CUSUM} is that the oncoSNP identifies only two regions with LOH and that the regions detected by oncoSNP are also detected by the proposed method. The largest of these regions coincides with that identified by oncoSNP and the largest number found by our methods justifies specificity values between 0.70 and 0.90.

The effect of choosing $m$ is not difficult to interpret and is clearly justified by the \autoref{comparizon_CUSUM}. Smaller values of $m$ provide a segmentation with smaller identified regions. The effect of $\delta$ is more subtle and can be seen in \autoref{comparizon_CUSUM} by a reduction on the first identified segment in panels (c) and (d). This happens because $\delta$ decreases the method's tolerance for the existence of points close to 0.5 within regions declared to contain LOH. Note that in the largest region with identified LOH there is a near-half observation for all segmentations built by our methods. This is because this observation is isolated within the segment and therefore its influence is not as strong as that in the first segment identified by the last two methods considered.

\section{Discussion}

The segmentation method we propose is conceptually simple in addition to being easily implemented.  A mixture of inflated betas models the BAF data allowing a fast model estimation procedure with the help of the EM algorithm. We segment the data set with the CUSUM technical which originated in the statistical process control and has optimal characteristics, resulting in a fast and accurate segmentation.

The great advantages of the method are that the estimation procedure can be performed in a small portion of the data set, result in a faster execution time in comparison to the methods that use HMM-based approaches. Also the built segmentation features can be adjusted by correctly selecting the parameters $m$, associated to the minimal length of a segment, and $\delta$, related to the level of tolerance to near-half observations inside LOH regions. The method is robust to mild levels of contaminations.

The fast performance characteristic of the proposed method is of great importance to the analysis of most recent array technology given the large amount of data resulting from application of said technologies. For example most recent SNP arrays can produce more than 2millions observations and whole-genome shotgun sequencing produce approximately 3billions observations, one for each base in the DNA chain. The sheer size of those numbers make clear the need for fast methods in the analysis of genetic related experiments. We believe that our proposed method is a advancement in this direction. 

\section*{Acknowledgements}
We want to thank FAPESP (grant 2013/00506-1) and CAPES for providing the resources for the realization of this project.

\bibliography{biblio_BAF.bib}

\end{document}